# The local spiral structure of the Milky Way

Ye Xu,[1]* Mark Reid,[2] Thomas Dame,[2] Karl Menten,[3] Nobuyuki Sakai,[4] Jingjing Li,[1,3] Andreas Brunthaler,[3] Luca Moscadelli,[5] Bo Zhang,[6] Xingwu Zheng[7]

The nature of the spiral structure of the Milky Way has long been debated. Only in the last decade have astronomers been able to accurately measure distances to a substantial number of high-mass star-forming regions, the classic tracers of spiral structure in galaxies. We report distance measurements at radio wavelengths using the Very Long Baseline Array for eight regions of massive star formation near the Local spiral arm of the Milky Way. Combined with previous measurements, these observations reveal that the Local Arm is larger than previously thought, and both its pitch angle and star formation rate are comparable to those of the Galaxy's major spiral arms, such as Sagittarius and Perseus. Toward the constellation Cygnus, sources in the Local Arm extend for a great distance along our line of sight and roughly along the solar orbit. Because of this orientation, these sources cluster both on the sky and in velocity to form the complex and long enigmatic Cygnus X region. We also identify a spur that branches between the Local and Sagittarius spiral arms.

## INTRODUCTION

The idea that the Milky Way is a spiral galaxy was proposed more than one and a half centuries ago (1). However, it was not until the 1950s that some spiral arm segments in the solar neighborhood were clearly identified (2, 3). Since then, many models have been proposed (4) and debated (5). Popular models (6, 7) suggest a grand design morphology with two- or four-armed spiral structure (8), such as M 51 or NGC 1232 (Fig. 1). Still, today, there is no general agreement on the number, locations, orientations, and properties of the Milky Way's spiral arms. The main impediments to determining spiral structure are the vast distances (up to about 60,000 ly) to stars across its extent and extinction by interstellar dust that precludes optical observations of distant stars in the Galactic plane. Recently, Very Long Baseline Interferometry (VLBI) at radio wavelengths has yielded near micro–arc second accurate position measurements to clouds of gas in regions of massive star formation, allowing estimation of distance through trigonometric parallax [astronomical "surveying" using Earth's orbit as a baseline (9)]. With the combined efforts of the U.S./European BeSSeL (Bar and Spiral Structure Legacy) Survey and the Japanese VERA (VLBI Exploration of Radio Astrometry) project, now more than 100 parallax measurements are locating large portions of spiral arms across the Milky Way (10, 11).

The Local Arm is the nearest spiral arm to the Sun. Because the nearby Orion stellar association lies within this arm, it has been called the "Orion Arm" or "Orion spur" (12).

Until recently, most available optical and radio data suggested that the Local Arm was a "spur" or secondary spiral feature [for example, (6)]. However, a large number of star-forming regions were recently measured to be in the Local Arm, many of them previously thought to be in the next more distant Perseus Arm. This suggested that the Local Arm is a major spiral structure [(13), hereafter *Paper I*]. Following *Paper*

*I*, we report eight new parallax measurements to stars with bright molecular maser emission at radio wavelengths. These provide a better understanding of the properties of the Local Arm, including defining its great extent, elucidating the nature of the well-studied Cygnus X region of massive star formation, and the discovery of a true spur connecting the Local and Sagittarius spiral arms.

## RESULTS

Spiral arms are best defined by the high-mass star-forming regions (HMSFRs) that form within them. We focused our study on 6.7-GHz methanol ($CH_3OH$) and 22-GHz water ($H_2O$) masers that are associated with well-known HMSFR sources, such as ultracompact HII regions and bright far-infrared sources. Following previous BeSSeL Survey studies of other spiral arms (11), we identify masers associated with the Local Arm based on their coincidence in Galactic longitude (l) and velocities in the Local Standard of Rest frame (V) with a large lane of emission at low absolute velocities that marks the Local Arm in CO and HI l-V diagrams. Table 1 presents the measured parallax distances for the eight new masers. Their spatial distribution in projection onto the Galactic plane is shown by the blue circles with white outlines in Fig. 2;

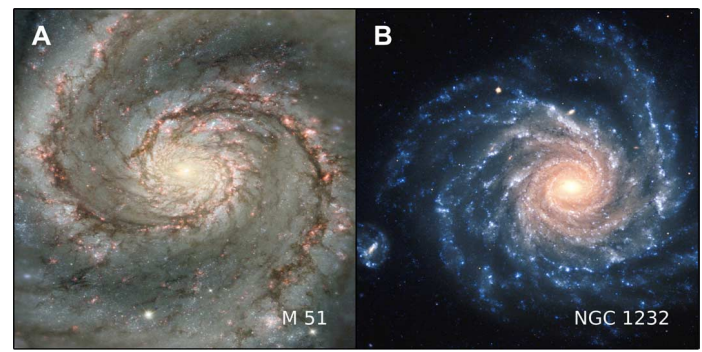

**Fig. 1. Spiral galaxies.** (**A**) M 51 [Credit: Hubble Space Telescope WFPC2]. (**B**) NGC 1232 [Credit: FORS, 8.2-meter Very Large Telescope Antu, European Southern Observatory].

[1]Purple Mountain Observatory, Chinese Academy of Sciences, Nanjing 210008, China. [2]Harvard-Smithsonian Center for Astrophysics, 60 Garden Street, Cambridge, MA 02138, USA. [3]Max-Planck-Institut für Radioastronomie, Auf dem Hügel 69, 53121 Bonn, Germany. [4]Mizusawa VLBI (Very Long Baseline Interferometry) Observatory, National Astronomical Observatory of Japan, Japan. [5]INAF (Istituto Nazionale di Astrofisica)–Osservatorio Astrofisico di Arcetri, Largo E. Fermi 5, 50125 Firenze, Italy. [6]Shanghai Astronomical Observatory, Chinese Academy of Sciences, Shanghai 200030, China. [7]Nanjing University, Nanjing 210093, China.
*Corresponding author. Email: xuye@pmo.ac.cn

**Table 1. Parallaxes and proper motions.** Column 2 is the parallax in milli–arc seconds. Column 3 is the parallax converted to distance in kiloparsec (1 pc ≈ 3.26 ly). Columns 4 and 5 are proper motions on the sky, eastward ($\mu_x = \mu_\alpha \cos\delta$) and northward ($\mu_y = \mu_\delta$) in units of milli–arc second year$^{-1}$, respectively.

| Maser | $\Pi$ | $D_\Pi$ | $\mu_x$ | $\mu_y$ |
|---|---|---|---|---|
| Name | (milli–arc second) | (kpc) | (milli–arc second year$^{-1}$) | (milli–arc second year$^{-1}$) |
| G054.10−00.08 | 0.231 ± 0.031 | $4.33^{+0.67}_{-0.51}$ | −3.13 ± 0.48 | −5.57 ± 0.48 |
| G058.77+00.64 | 0.299 ± 0.040 | $3.34^{+0.52}_{-0.39}$ | −2.70 ± 0.10 | −6.10 ± 0.21 |
| G059.47+00.18 | 0.535 ± 0.024 | $1.87^{+0.09}_{-0.08}$ | −1.83 ± 1.12 | −6.60 ± 1.12 |
| G059.83+00.67 | 0.253 ± 0.024 | $3.95^{+0.42}_{-0.34}$ | −2.92 ± 0.07 | −6.03 ± 0.05 |
| G071.52−00.38 | 0.277 ± 0.013 | $3.61^{+0.18}_{-0.16}$ | −2.48 ± 0.04 | −4.97 ± 0.07 |
| G108.18+05.51 | 1.101 ± 0.033 | $0.91^{+0.03}_{-0.03}$ | +0.16 ± 0.09 | −2.17 ± 0.35 |
| G109.87+02.11 | 1.208 ± 0.025 | $0.83^{+0.02}_{-0.02}$ | −1.03 ± 0.10 | −2.62 ± 0.27 |
| G213.70−12.60 | 1.166 ± 0.021 | $0.86^{+0.01}_{-0.02}$ | −1.25 ± 0.09 | +2.44 ± 0.28 |

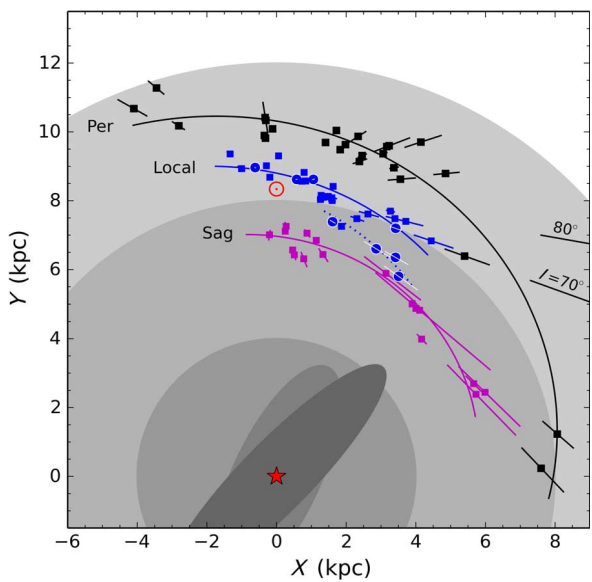

**Fig. 2. Location of HMSFRs determined by trigonometric parallax.** Sources presented in this paper are indicated with the blue circles with white outlines, and those from *Paper I* with blue squares. The parallax data in the Perseus (black squares) and Sagittarius (magenta squares) arms are also presented. The blue solid and dot lines are log spiral fits to the sources in the Local Arm and in the spur, respectively. They have pitch angles of 11.6° ± 1.8° and 18.3° ± 5.9°. The Galactic center (red star) is at (0,0) and the Sun (red Sun symbol) is at (0,8.34). Distance error bars (1σ) are indicated, but many are smaller than the symbols. The background gray disks provide scale, with radii corresponding to round numbers for the Galactic bar region (≈4 kpc), the solar circle (≈8 kpc), and corotation of the spiral pattern and Galactic orbits (≈12 kpc). The short Cosmic Background Explorer "boxy-bar" and the "long" bar (*19–21*) are indicated with shaded ellipses.

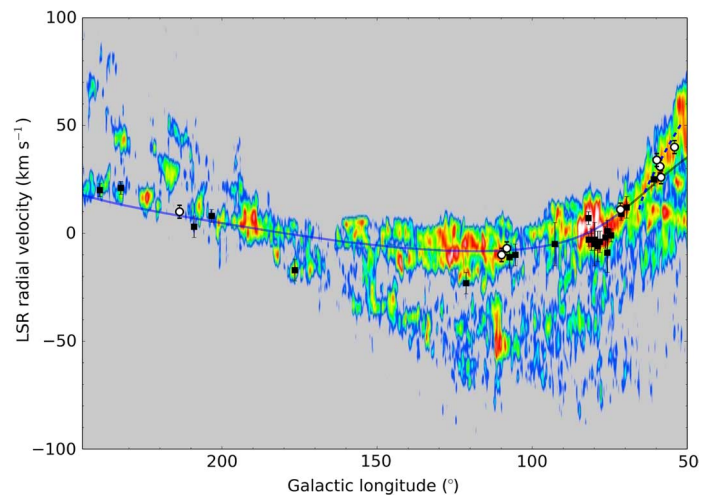

**Fig. 3. Location of HMSFRs with parallaxes superposed on a CO *l-V* diagram from the CfA 1.2-m survey (*22*).** Sources presented in this paper are indicated with white circles with black outlines, and others with black squares. Velocity error bars (1σ) are indicated. The solid line is a log spiral fit to the sources in the Local Arm (blue for *l* > 70°, black for an extrapolation at *l* < 70°), whereas the dotted line corresponds to the spur indicated in Fig. 2. LSR, Local Standard of Rest.

the blue squares are the sources from *Paper I*. For Galactic context, we also show HMSFRs in the surrounding Sagittarius and Perseus arms (*11*). Almost all of the sources in the Local Arm have a distance accuracy of better than ±10%, and half are better than ±5%, ensuring that the spiral structure near the Sun can be mapped with unprecedented accuracy.

## DISCUSSION

Our newly measured sources at *l* > 70° strengthen the main conclusion of *Paper I* that the Local Arm is quite long and has a modest pitch angle and an abundant star formation. The addition of these sources improves the accuracy and robustness of our pitch angle estimate (blue solid curve) from 10.1° ± 2.7° to 11.6° ± 1.8°. We now know that the density of HMSFRs in the Local Arm is comparable to that of other major arms and that it stretches for >20,000 ly, almost reaching the Perseus Arm (*11* and this paper). With standard density-wave theory for grand-design morphologies, it would be difficult to explain this large spiral arm segment located between the Sagittarius and Perseus arms, owing to its narrow spacing (*14*). This suggests that the Milky Way does not have a pure grand design. Also, recent large numerical simulations suggest that standard density-wave theory may not explain spiral structure in galaxies like the Milky Way (15).

The four new parallax-measured sources at $l < 70°$ do not follow the main arc of the Local Arm. Instead, these sources, as well as G059.78+00.06 and ON 1, branch off and curve inward in the Milky Way. As the dotted line in Fig. 2 suggests, these sources trace what appears to be a high-inclination spur bridging the Local Arm to the Sagittarius Arm near $l ≈ 50°$. Additional evidence for this spur is provided by large-scale molecular emission from CO, as shown in the $l$-$V$ diagram in Fig. 3. Assuming circular motion and a flat rotation curve for the Milky Way, we can calculate the velocity $V$ of Galactic objects using $V = \Theta_0(R_0/R − 1)$ sin$l$, where $R$ is the Galactocentric radius and $R_0$ and $\Theta_0$ are the distance to the Galactic center and the circular rotation speed at the Sun, respectively (11). In this way, the spatial locations of the Local Arm and spur sources in Fig. 2 can be transformed into $l$-$V$ space, as shown in Fig. 3. At $l < 70°$, velocities from the spiral fit fall significantly below some of the spur sources (G054.10−00.08 and G059.83+00.67). Note also that they follow a nearly continuous lane of CO emission that runs from the intense Cygnus X region near $l ≈ 80°$, almost certainly part of the Local Arm, to the Sagittarius Arm near $l ≈ 50°$. This lane has received little attention in the past because it does not correspond with any of the major spiral arm features of the inner Galaxy [for example, (16)].

At distances beyond the branch point of the spur, the HMSFRs in the Local Arm align almost linearly toward $l \sim 80°$ (Fig. 2). Because these sources straddle the solar orbit, they pile up at a point in the $l$-$V$ diagram (Fig. 3) near zero velocity. CO emission associated with the arm likewise piles up to form the complex Cygnus X region. Whether this region is a superposition of HMSFRs strung along the line of sight [for example, (17)] or mainly a single giant molecular complex at one distance [for example, (18)] has been debated for decades. Our results strongly support the former view, with the sources toward Cygnus X extending more than 13,000 ly and nearly reaching the Perseus Arm.

## MATERIALS AND METHODS

Observations with 7-hour tracks for 6.7-GHz $CH_3OH$ and 22-GHz $H_2O$ masers toward star-forming regions were performed with the National Radio Astronomy Observatory Very Long Baseline Array (VLBA), under programs BR149 and BR198 as part of the BeSSeL Survey. These are the first 6.7-GHz $CH_3OH$ maser parallax measurements with the VLBA.

For 6.7-GHz $CH_3OH$ masers, four observing epochs were selected to optimally sample the peaks of the sinusoidal parallax signature in right ascension over 1 year, maximizing the sensitivity of parallax detection and ensuring that the parallax and proper motion signatures are uncorrelated. For 22-GHz $H_2O$ masers, six epochs were observed in 1 year to allow a parallax measurement with less than one full year of data, because water maser spots can have lifetimes shorter than a year. Details about the observations and data analysis are available in the Supplementary Materials.

## SUPPLEMENTARY MATERIALS

Supplementary material for this article is available at http://advances.sciencemag.org/cgi/content/full/2/9/e1600878/DC1

Supplementary Text

table S1. Details of the epochs observed.

table S2. Positions and source properties of the reference maser spots and the background sources for the first epoch.

table S3. Parallaxes and proper motions.

table S4. Parallax and proper motion measurements.

fig. S1. Parallax and proper motion data and fits for G054.10−00.08.

fig. S2. Parallax and proper motion data and fits for G058.77+00.64.

fig. S3. Parallax and proper motion data and fits for G059.47−00.18.

fig. S4. Parallax and proper motion data and fits for G059.83+00.67.

fig. S5. Parallax and proper motion data and fits for G071.52−00.38.

fig. S6. Parallax and proper motion data and fits for G108.18+05.51.

fig. S7. Parallax and proper motion data and fits for G109.87+02.11.

fig. S8. Parallax and proper motion data and fits for G213.70−12.60.

References (23–39)

**Acknowledgments:** We thank the anonymous referee for many useful comments that have improved the paper. **Funding:** This work was supported by the National Natural Science Foundation of China (grant nos. 11133008 and 11233007), the Strategic Priority Research Program of the Chinese Academy of Sciences (grant no. XDB09010300), and the Key Laboratory of Radio Astronomy. **Author contributions:** Y.X., M.R., T.D., and K.M. contributed to science discussions. Y.X., M.R., and T.D. wrote the manuscript. Y.X., J.L., and A.B. performed data analysis. N.S., L.M., B.Z., and X.Z. improved the manuscript. All authors have read and approved the manuscript. **Competing interests:** The authors declare that they have no competing interests. **Data and materials availability:** All data needed to evaluate the conclusions in the paper are present in the paper and/or the Supplementary Materials. Additional data related to this paper may be requested from the authors.

Submitted 23 April 2016
Accepted 18 August 2016
Published 28 September 2016
10.1126/sciadv.1600878

**Citation:** Y. Xu, M. J. Reid, T. Dame, K. Menten, N. Sakai, J. Li, A. Brunthaler, L. Moscadelli, B. Zhang, X. Zheng, The local spiral structure of the Milky Way. *Sci. Adv.* **2**, e1600878 (2016).

**The following resources related to this article are available online at
http://advances.sciencemag.org. (This information is current as of October 1, 2016):**

**Updated information and services,** including high-resolution figures, can be found in the
online version of this article at:
http://advances.sciencemag.org/content/2/9/e1600878.full

**Supporting Online Material** can be found at:
http://advances.sciencemag.org/content/suppl/2016/09/26/2.9.e1600878.DC1

This article **cites 36 articles,** 5 of which you can access for free at:
http://advances.sciencemag.org/content/2/9/e1600878#BIBL

# Supplementary Materials for

## The local spiral structure of the Milky Way

Ye Xu, Mark Reid, Thomas Dame, Karl Menten, Nobuyuki Sakai, Jingjing Li, Andreas Brunthaler,
Luca Moscadelli, Bo Zhang, Xingwu Zheng

**This PDF file includes:**

**Supplementary Text**

**Observations and Analysis**

Our observational setup is similar to that of the previous work (*23*) except for two details. First, in the phase-referenced observations, we employed four adjacent, dual circularly polarized bands of 16 MHz, which were correlated to give 1000 and 2000 spectral channels for BR149 and BR198, respectively. This corresponds to a velocity resolution of 0.36 km s$^{-1}$ for the 6.7-GHz CH$_3$OH masers for the former, and 0.18 km s$^{-1}$ and 0.11 km s$^{-1}$ for the 6.7-GHz CH$_3$OH and 22-GHz H$_2$O masers for the latter. Second, in order to remove uncompensated interferometric delays introduced by the Earth's troposphere and ionosphere, dual-frequency "geodetic block" observations were used for the 6.7-GHz CH$_3$OH masers. These data were taken in left circular polarization with four 16 MHz bands that spanned ≈ 480 MHz of bandwidth from 4.116 to 4.592 GHz and from 7.094 to 7.570 GHz; the bands were spaced in a minimum redundancy configuration with frequencies offset by 4, 80, 320 and 480 MHz so as to uniformly sample frequency differences.

Correlated data from both groups of bands were divided into separate data files, corrected for ionospheric delays using total electron content models, and residual multi-band delays were determined for all sources. Dispersive "ionosphere-only" delays were formed by differencing the delays at 4.3 and 7.3 GHz, and then scaled to 6.7 GHz. Non-dispersive (tropospheric) delays were obtained by removing the dispersive component from the total observed delay at 7.3 GHz. Then non-dispersive delays were modeled as owing to a vertical atmospheric delay and a clock offset and drift rate at each antenna. Since the clock terms cancel in the pure dispersive delays, these were modeled as a vertical ionospheric delay.

For parallax measurements, our selection criteria included a minimum flux density of 3 Jy for 6.7-GHz methanol masers and 8 Jy for 22-GHz water masers, and we required at least 2 nearby background sources with separations less than 3 degrees. The background source selection excluded less than 10% of strong masers. The masers were selected from pre-existing catalogs, whose completeness is not well documented, but is likely to be at least 80% complete at our flux density cutoffs. So, we estimate that we have measured parallaxes for more than 70% of high mass star forming regions with 6.7 GHz stronger than 3 Jy and 22 GHz masers stronger than 8 Jy. Background sources were selected from the VCS2 and VCS3 catalogs (*24, 25*) and the BeSSeL calibration survey (*26, 27*). Except for G059.83+00.67, where the extragalactic radio sources were used as the phase reference, a clean maser channel was used for other sources. Table S2 shows the positions and source properties for the masers used for parallax measurements and the corresponding background sources, as well as other observational parameters.

G054.10−00.08 & G059.47−00.18 were observed at 22-GHz (H$_2$O masers) and the other six sources at 6.7-GHz (CH$_3$OH masers). Parallax and proper motion were estimated following techniques described in previous work (*28, Paper I*). $H_2O$ masers are found in outflows with speeds of $10 - 100$ km s$^{-1}$, and the maser spots are not always distributed uniformly around their central exciting star. This can limit the estimated accuracy of the proper motion of the exciting star. Since the two $H_2O$ maser sources have few spots and a simple narrow maser spectrum, the motions of their central stars were determined by averaging the persistent spots used for parallax fitting, and then assigning a proper motion uncertainty of 10 km s$^{-1}$ at the measured distance. Unlike $H_2O$ masers, $CH_3OH$ masers move slowly, typically a few km s$^{-1}$ (*29*), so we expect little problem in estimating the accuracy of the proper motions of the underlying stars.

**table S1. Details of the epochs observed.**

| Code | Freq.(GHz) | Source | Epoch 1 | Epoch 2 | Epoch 3 | Epoch 4 | Epoch 5 | Epoch 6 | Epoch 7 |
|------|-----------|--------|---------|---------|---------|---------|---------|---------|---------|
| BR149R | 6.7 | G059.83+00.67 | 12 Nov 18 | 13 Apr 12 | 13 Oct 25 | 13 Nov 02 | 14 Apr 29 | | |
| BR149S | 6.7 | G109.87+02.11 | 12 Dec 08 | 13 May 19 | 13 Jun 24 | 13 Nov 24 | | | |
| BR149U | 6.7 | G108.18+05.51 | 12 Dec 03 | 13 May 12 | 13 Jun 06 | 13 Nov 23 | | | |
| BR198S | 22 | G054.10−00.08 | 13 Nov 07 | 14 Apr 18 | 14 Jul 25 | 14 Sep 22 | 14 Nov 09 | 14 Dec 18 | 15 May 03 |
| BR198T | 6.7 | G058.77+00.64 | 13 Nov 09 | 14 Apr 20 | 14 Oct 30 | 14 Nov 10 | 15 May 04 | | |
| | | G071.52−00.38 | | 14 Apr 20 | 14 Oct 30 | 14 Nov 10 | 15 May 04 | | |
| BR198U | 22 | G059.47−00.18 | 13 Nov 11 | 14 May 20 | 14 Aug 04 | 14 Sep 21 | 14 Nov 11 | 14 Dec 19 | 15 May 07 |
| BR198V | 6.7 | G213.70−12.60 | 13 Sep 24 | 14 Mar 20 | 14 Sep 25 | 14 Oct 17 | 15 Mar 26 | | |

**table S2. Positions and source properties of the reference maser spots and the background sources for the first epoch.** Columns 2 and 3 are Right Ascension (R.A.) and Declination (Dec.), respectively. Column 4 - 6 are the angular separation between the maser and the calibrator, the peak flux density and the velocity of the reference maser spot. Column 7 is the interferometer restoring beam. Column 8 lists the phase-reference source.

| Source | R.A. (J2000) ($^{h\ m\ s}$) | Dec. (J2000) ($^{\circ\ \prime\ \prime\prime}$) | $\phi$ ($^{\circ}$) | Brightness (Jy beam$^{-1}$) | $V$ (km s$^{-1}$) | Beam (mas×mas, $^{\circ}$) | Phase-reference Source |
|---|---|---|---|---|---|---|---|
| G054.10−0.08 | 19 31 48.7978 | +18 42 57.096 | | 5.4 | 37.7 | 1.3×0.7 @ −16 | Y |
| J1928+1859 | 19 28 42.7191 | +18 59 24.563 | 0.8 | 0.02 | | 1.3×0.7 @ −18 | |
| J1935+2031 | 19 35 10.47290 | +20 31 54.1540 | 2.0 | 0.17 | | 2.1×1.0 @ −46 | |
| G058.77+00.64 | 19 38 49.1269 | +23 08 40.205 | | 2.7 | 33.4 | 6.5×2.4 @ −33 | Y |
| J1936+2246 | 19 36 29.3047 | +22 46 25.851 | 0.7 | 0.06 | | 7.2×2.8 @ −34 | |
| J1936+2357 | 19 36 00.9250 | +23 57 31.973 | 1.0 | 0.10 | | 6.8×3.0 @ −37 | |
| G059.47−00.18 | 19 43 28.3504 | +23 20 42.522 | | 20.8 | 23.1 | 1.2×0.7 @ −14 | Y |
| J1941+2307 | 19 41 55.1115 | +23 07 56.521 | 0.4 | 0.01 | | 1.2×0.7 @ −14 | |
| J1946+2300 | 19 46 06.25140 | +23 00 04.4147 | 0.7 | 0.08 | | 1.2×0.7 @ −14 | |
| J1946+2255 | 19 46 22.2995 | +22 55 24.647 | 0.9 | 0.05 | | 1.2×0.7 @ −14 | |
| G059.83+00.67 | 19 40 59.2938 | +24 04 44.177 | | 2.4 | 38.0 | 4.6×1.6 @ −16 | Y |
| J1936+2357 | 19 36 00.9250 | +23 57 31.973 | 1.1 | 0.12 | | 3.0×1.4 @ −7 | Y |
| J1946+2418 | 19 46 19.9669 | +24 18 56.789 | 1.2 | 0.02 | | 3.0×1.4 @ −6 | |
| G071.52−00.38 | 20 12 57.8943 | +33 30 27.083 | | 3.1 | 10.1 | 7.4×4.4 @ +30 | Y |
| J2015+3410 | 20 15 28.83190 | +34 10 39.4100 | 0.8 | 0.23 | | 7.3×4.4 @ +23 | |
| G108.18+05.51 | 22 28 51.4117 | +64 13 41.201 | | 1.4 | -12.8 | 5.4×2.4 @ −20 | Y |
| J2237+6408 | 22 37 30.6578 | +64 08 53.942 | 0.9 | 0.05 | | 6.7×2.4 @ −18 | |
| J2223+6249 | 22 23 18.09660 | +62 49 33.8055 | 1.5 | 0.08 | | 5.8×2.4 @ −30 | |
| J2231+6218 | 22 31 43.2609 | +62 18 51.734 | 1.9 | 0.03 | | 6.4×2.7 @ −18 | |
| G109.87+02.11 | 22 56 18.0559 | +62 01 49.562 | | 17.6 | -2.2 | 5.5×3.4 @ −32 | Y |
| J2254+6209 | 22 54 25.2927 | +62 09 38.725 | 0.3 | 0.03 | | 5.3×3.4 @ −30 | |
| J2243+6055 | 22 43 00.9439 | +60 55 44.226 | 1.9 | 0.06 | | 5.7×3.4 @ −35 | |
| J2302+6405 | 23 02 41.31500 | +64 05 52.8485 | 2.2 | 0.11 | | 5.2×3.4 @ −31 | |
| G213.70−12.60 | 06 07 47.8632 | −06 22 56.511 | | 25.2 | 12.3 | 4.3×1.5 @ −17 | Y |
| J0603-0645 | 06 03 38.05676 | −06 45 44.8646 | 1.1 | 0.06 | | 4.5×2.6 @ −40 | |
| J0606-0724 | 06 06 43.54629 | −07 24 30.2323 | 1.1 | 0.15 | | 4.2×2.2 @ −13 | |
| J0605-0759 | 06 05 26.18727 | −07 59 31.7893 | 1.7 | 0.04 | | 4.3×2.6 @ −37 | |
| J0607-0834 | 06 07 59.69924 | −08 34 49.9782 | 2.2 | 1.60 | | 4.1×2.2 @ −14 | |

**table S3. Parallaxes and proper motions.** Columns 2 and 3 are Galactic longitude and latitude, respectively. Columns 4 − 6 are the parallax and proper motion in the eastward ($\mu_x = \mu_\alpha \cos\delta$ ) and northward directions ($\mu_y = \mu_\delta$), respectively. Column 7 is the LSR velocity. The sources with $l < 70°$ consist of the spur, while others belong to the Local Arm.

| Source | $l$ | $b$ | Parallax | $\mu_x$ | $\mu_y$ | $V$ | Ref. |
|---|---|---|---|---|---|---|---|
| | (°) | (°) | (mas) | (mas y$^{-1}$) | (mas y$^{-1}$) | (km s$^{-1}$) | |
| G054.10−0.08 | 54.1098 | −0.0812 | 0.231 ± 0.031 | −3.13 ± 0.48 | −5.57 ± 0.48 | 40 ± 3 | (*30, this paper*) |
| G058.77+00.64 | 58.7748 | +0.6442 | 0.299 ± 0.040 | −2.70 ± 0.10 | −6.10 ± 0.21 | 33 ± 3 | (*30, this paper*) |
| G059.47−00.18 | 59.4779 | −0.1856 | 0.535 ± 0.024 | −1.83 ± 1.12 | −6.60 ± 1.12 | 26 ± 3 | (*31, this paper*) |
| G059.78+00.06 | 59.7829 | +0.0647 | 0.463 ± 0.020 | −1.65 ± 0.03 | −5.12 ± 0.08 | 25 ± 3 | (*13*) |
| G059.83+00.67 | 59.8328 | +0.6721 | 0.253 ± 0.024 | −2.92 ± 0.07 | −6.03 ± 0.05 | 34 ± 3 | (*30, this paper*) |
| ON 1 | 69.5402 | −0.9755 | 0.406 ± 0.031 | −3.19 ± 0.37 | −5.22 ± 0.25 | 12 ± 5 | (*13*) |
| IRAS 20056+3350 | 71.3126 | +0.8290 | 0.213 ± 0.026 | −2.62 ± 0.33 | −5.65 ± 0.52 | 9 ± 3 | (*32*) |
| G071.52−00.38 | 71.5220 | −0.3851 | 0.277 ± 0.013 | −2.48 ± 0.04 | −4.97 ± 0.07 | 11 ± 3 | (*33, this paper*) |
| IRAS 20143+3634 | 74.5695 | +0.8459 | 0.367 ± 0.037 | −2.99 ± 0.16 | −4.37 ± 0.43 | −1 ± 3 | (*34, 35*) |
| G075.76+00.33 | 75.7610 | +0.3400 | 0.285 ± 0.022 | −3.08 ± 0.06 | −4.56 ± 0.08 | −9 ± 9 | (*13*) |
| G075.78+00.34/ON 2N | 75.7821 | +0.3428 | 0.271 ± 0.022 | −2.79 ± 0.10 | −4.69 ± 0.12 | 1 ± 5 | (*13*) |
| G076.38−00.61 | 76.3809 | −0.6177 | 0.770 ± 0.053 | −3.73 ± 3.00 | −3.84 ± 3.00 | −2 ± 5 | (*13*) |
| IRAS 20126+4104 | 78.1224 | +3.6328 | 0.610 ± 0.020 | −4.14 ± 0.13 | −4.14 ± 0.13 | −4 ± 5 | (*13*) |
| AFGL 2591 | 78.8867 | +0.7090 | 0.300 ± 0.010 | −1.20 ± 0.32 | −4.80 ± 0.12 | −6 ± 7 | (*13*) |
| IRAS 20290+4052 | 79.7358 | +0.9905 | 0.737 ± 0.062 | −2.84 ± 0.09 | −4.14 ± 0.54 | −3 ± 5 | (*13*) |
| G079.87+01.17 | 79.8769 | +1.1770 | 0.620 ± 0.027 | −3.23 ± 1.31 | −5.19 ± 1.31 | −5 ± 10 | (*13*) |
| NML Cyg | 80.7984 | −1.9209 | 0.620 ± 0.047 | −1.55 ± 0.42 | −4.59 ± 0.41 | −3 ± 3 | (*13*) |
| DR 20 | 80.8615 | +0.3834 | 0.687 ± 0.038 | −3.29 ± 0.13 | −4.83 ± 0.26 | −3 ± 5 | (*13*) |
| DR 21 | 81.7524 | +0.5908 | 0.666 ± 0.035 | −2.84 ± 0.15 | −3.80 ± 0.22 | −3 ± 3 | (*13*) |
| W 75N | 81.8713 | +0.7807 | 0.772 ± 0.042 | −1.97 ± 0.10 | −4.16 ± 0.15 | 7 ± 3 | (*13*) |
| G092.67+03.07 | 92.6712 | +3.0712 | 0.613 ± 0.020 | −0.69 ± 0.26 | −2.25 ± 0.33 | −5 ± 10 | (*13*) |
| G105.41+09.87 | 105.4154 | +9.8783 | 1.129 ± 0.063 | −0.21 ± 2.38 | −5.49 ± 2.38 | −10 ± 5 | (*13*) |
| IRAS 22198+6336 | 107.2982 | +5.6398 | 1.309 ± 0.047 | −2.47 ± 0.21 | +0.26 ± 0.40 | −11 ± 5 | (*13*) |
| G108.18+05.51/L 1206 | 108.1844 | +5.5189 | 1.101 ± 0.033 | +0.16 ± 0.09 | −2.17 ± 0.35 | −11 ± 3 | (*36, this paper*) |
| G109.87+02.11/Cep A | 109.8711 | +2.1144 | 1.208 ± 0.025 | −1.03 ± 0.10 | −2.62 ± 0.27 | −10 ± 3 | (*37, this paper*) |
| L 1287 | 121.2978 | +0.6588 | 1.077 ± 0.039 | −0.86 ± 0.11 | −2.29 ± 0.56 | −23 ± 5 | (*13*) |
| G176.51+00.20 | 176.5156 | +0.2009 | 1.038 ± 0.021 | +1.84 ± 1.00 | −5.86 ± 1.00 | −17 ± 5 | (*32*) |
| NGC 2264 | 203.3209 | +2.0530 | 1.356 ± 0.098 | −0.96 ± 0.58 | −6.05 ± 3.06 | 8 ± 3 | (*38, 39*) |
| Orion | 209.0071 | −19.3854 | 2.408 ± 0.033 | +3.30 ± 1.50 | +0.10 ± 1.50 | 3 ± 5 | (*13*) |
| G213.70−12.60/Mon R2 | 213.7049 | −12.5969 | 1.166 ± 0.021 | −1.25 ± 0.09 | +2.44 ± 0.28 | 10 ± 3 | (*37, this paper*) |
| G232.62+00.99 | 232.6205 | +0.9956 | 0.596 ± 0.035 | −2.17 ± 0.06 | +2.09 ± 0.46 | 21 ± 3 | (*13*) |
| VY CMa | 239.3526 | −5.0655 | 0.855 ± 0.080 | −2.80 ± 0.20 | +2.60 ± 0.20 | 20 ± 3 | (*13*) |

## Parallax and Proper Motion Fitting Details

Parallax fits are based on spots persisting over four or more epochs, as shown in table S4. The parallax uncertainties in the table are formal fitting uncertainties (scaled to give a chi-squared-per-degree-of-freedom of unity) and then multiplied by sqrt(N), where N is the number of maser spots used, in order to conservatively allow for correlated systematic error from uncompensated atmospheric delays between the masers and a QSO.

Figure S1 to S8 show the maser spot positions (relative to the background sources) as a function of time and the parallax fits.

### table S4. Parallax and proper motion measurements.

| Background Source | $V_{LSR}$ (km s$^{-1}$) | Detected epochs | Parallax (mas) | $\mu_x$ (mas yr$^{-1}$) | $\mu_y$ (mas yr$^{-1}$) |
|---|---|---|---|---|---|
| **G054.10−00.08** | | | | | |
| J1928+1859 | 36.2 | 1111101 | 0.194±0.016 | −3.18±0.03 | −5.76±0.08 |
| | 36.2 | 1111101 | 0.203±0.026 | −3.38±0.05 | −5.86±0.08 |
| | 37.7 | 1111101 | 0.261±0.056 | −3.06±0.11 | −5.24±0.12 |
| | Combined fit | | 0.217±0.022 | | |
| | Average | | | −3.17±0.06 | −5.53±0.09 |
| J1935+2031 | 36.2 | 1111101 | 0.231±0.037 | −3.11±0.07 | −5.84±0.13 |
| | 36.2 | 1111101 | 0.232±0.043 | −3.30±0.08 | −5.93±0.10 |
| | 37.7 | 1111101 | 0.289±0.061 | −2.98±0.12 | −5.32±0.14 |
| | Combined fit | | 0.248±0.028 | | |
| | Average | | | −3.09±0.09 | −5.60±0.12 |
| | **Combined fit** | | **0.231±0.018** | | |
| | **Average** | | | **−3.13±0.09** | **−5.57±0.12** |
| **G058.77+00.64** | | | | | |
| J1936+2246 | 33.4 | 11111 | 0.308±0.078 | −2.73±0.14 | −6.07±0.26 |
| | 34.8 | 11111 | 0.316±0.065 | −2.78±0.12 | −6.10±0.26 |
| | 35.5 | 11111 | 0.327±0.078 | −2.85±0.14 | −6.08±0.26 |
| | Combined fit | | 0.317±0.040 | | |
| | Average | | | −2.79±0.13 | −6.08±0.26 |
| J1936+2357 | 33.4 | 11111 | 0.273±0.031 | −2.54±0.06 | −6.09±0.14 |
| | 34.8 | 11111 | 0.290±0.029 | −2.60±0.05 | −6.13±0.15 |
| | 35.5 | 11111 | 0.292±0.033 | −2.67±0.06 | −6.11±0.15 |
| | Combined fit | | 0.285±0.017 | | |
| | Average | | | −2.60±0.06 | −6.11±0.15 |
| | **Combined fit** | | **0.299±0.023** | | |
| | **Average** | | | **−2.70±0.10** | **−6.10±0.21** |
| **G059.47−00.18** | | | | | |
| J1941+2307 | 23.1 | 1111111 | 0.593±0.033 | −1.98±0.06 | −6.42±0.07 |
| | 24.6 | 1111111 | 0.575±0.032 | −1.79±0.05 | −6.85±0.08 |
| | Combined fit | | 0.585±0.023 | | |
| | Average | | | −1.89±0.06 | −6.64±0.08 |
| J1946+2300 | 23.1 | 1111111 | 0.509±0.060 | −1.87±0.13 | −6.36±0.10 |
| | 24.6 | 1111111 | 0.510±0.059 | −1.70±0.13 | −6.80±0.10 |
| | Combined fit | | 0.510±0.041 | | |
| | Average | | | −1.79±0.13 | −6.58±0.10 |
| J1946+2255 | 23.1 | 1111111 | 0.481±0.037 | −1.88±0.10 | −6.35±0.05 |
| | 24.6 | 1111111 | 0.500±0.038 | −1.72±0.08 | −6.79±0.06 |
| | Combined fit | | 0.494±0.026 | | |
| | Average | | | −1.80±0.09 | −6.57±0.06 |
| | **Combined fit** | | **0.535±0.017** | | |
| | **Average** | | | **−1.83±0.09** | **−6.60±0.08** |
| **G059.83+00.67** | | | | | |
| J1936+2357 | 37.6 | 11111 | 0.294±0.030 | −2.85±0.06 | −5.90±0.10 |
| | 38.0 | 11111 | 0.279±0.033 | −2.90±0.06 | −5.96±0.05 |
| | 38.4 | 11111 | 0.265±0.024 | −2.91±0.05 | −5.99±0.03 |
| | 38.7 | 11111 | 0.259±0.022 | −2.91±0.05 | −5.96±0.05 |
| | Combined fit | | 0.277±0.014 | | |
| | Average | | | −2.89±0.06 | −5.95±0.06 |
| J1946+2418 | 37.6 | 11111 | 0.244±0.037 | −2.89±0.07 | −6.04±0.07 |
| | 38.0 | 11111 | 0.249±0.033 | −2.95±0.07 | −6.11±0.04 |
| | 38.4 | 11111 | 0.258±0.034 | −2.97±0.08 | −6.15±0.03 |
| | 38.7 | 11111 | 0.247±0.032 | −2.96±0.08 | −6.13±0.03 |
| | Combined fit | | 0.238±0.016 | | |
| | Average | | | −2.94±0.08 | −6.11±0.04 |
| | **Combined fit** | | **0.253±0.012** | | |
| | **Average** | | | **−2.92±0.07** | **−6.03±0.05** |
| **G071.52−00.38** | | | | | |
| J2015+3410 | 10.1 | 01111 | 0.273±0.017 | −2.35±0.05 | −5.04±0.06 |
| | 11.0 | 01111 | 0.279±0.010 | −2.60±0.03 | −4.89±0.07 |
| | **Combined fit** | | **0.277±0.009** | | |
| | **Average** | | | **−2.48±0.04** | **−4.97±0.07** |
| **G108.18+05.51** | | | | | |
| J2237+6408 | -12.8 | 1111 | 1.026±0.034 | +0.09±0.09 | −2.01±0.44 |
| | -13.2 | 1111 | 1.073±0.050 | +0.07±0.13 | −1.85±0.28 |
| | Combined fit | | 1.050±0.048 | | |
| | Average | | | +0.08±0.11 | −1.93±0.36 |
| J2223+6249 | -12.8 | 1111 | 0.944±0.028 | −0.06±0.07 | −2.13±0.51 |
| | -13.2 | 1111 | 0.995±0.010 | −0.08±0.02 | −1.96±0.20 |
| | Combined fit | | 0.970±0.015 | | |
| | Average | | | −0.07±0.05 | −2.05±0.36 |
| J2231+6218 | -12.8 | 1111 | 1.123±0.043 | +0.49±0.12 | −2.60±0.51 |
| | -13.2 | 1111 | 1.169±0.028 | +0.47±0.07 | −2.44±0.13 |
| | Combined fit | | 1.146±0.040 | | |
| | Average | | | +0.48±0.10 | −2.52±0.32 |
| | **Combined fit** | | **1.101±0.023** | | |
| | **Average** | | | **0.16±0.09** | **−2.17±0.35** |
| **G109.87+02.11** | | | | | |
| J2254+6209 | -2.2 | 1111 | 1.205±0.055 | −1.05±0.14 | −2.50±0.34 |
| J2243+6055 | -2.2 | 1111 | 1.166±0.020 | −0.94±0.04 | −2.95±0.28 |
| J2302+6405 | -2.2 | 1111 | 1.256±0.045 | −1.09±0.12 | −2.42±0.20 |
| | **Combined fit** | | **1.208±0.025** | | |
| | **Average** | | | **−1.03±0.10** | **−2.62±0.27** |
| **G213.70−12.60** | | | | | |
| J0603-0645 | 10.1 | 01111 | 1.165±0.014 | −1.57±0.04 | +2.15±0.40 |
| | 12.3 | 01111 | 1.091±0.030 | −1.04±0.08 | +3.13±0.29 |
| | 13.4 | 01111 | 1.148±0.035 | −1.34±0.09 | +2.87±0.35 |
| | Combined fit | | 1.134±0.020 | | |
| | Average | | | −1.32±0.07 | +2.72±0.35 |
| J0606-0724 | 10.1 | 01111 | 1.180±0.059 | −1.48±0.16 | +1.64±0.36 |
| | 12.3 | 01111 | 1.107±0.027 | −0.95±0.07 | +2.62±0.22 |
| | 13.4 | 01111 | 1.163±0.023 | −1.25±0.06 | +2.36±0.31 |
| | Combined fit | | 1.150±0.025 | | |
| | Average | | | −1.23±0.10 | +2.21±0.30 |
| J0605-0759 | 10.1 | 01111 | 1.189±0.029 | −1.55±0.08 | +1.76±0.29 |
| | 12.3 | 01111 | 1.113±0.012 | −1.02±0.03 | +2.74±0.04 |
| | 13.4 | 01111 | 1.170±0.014 | −1.32±0.04 | +2.48±0.25 |
| | Combined fit | | 1.156±0.016 | | |
| | Average | | | −1.30±0.05 | +2.33±0.19 |
| J0607-0834 | 10.1 | 01111 | 1.253±0.076 | −1.41±0.20 | +1.92±0.36 |
| | 12.3 | 01111 | 1.181±0.044 | −0.88±0.12 | +2.90±0.18 |
| | 13.4 | 01111 | 1.236±0.040 | −1.18±0.11 | +2.65±0.32 |
| | Combined fit | | 1.223±0.032 | | |
| | Average | | | −1.16±0.14 | +2.49±0.29 |
| | **Combined fit** | | **1.166±0.012** | | |
| | **Average** | | | **−1.25±0.09** | **+2.44±0.28** |

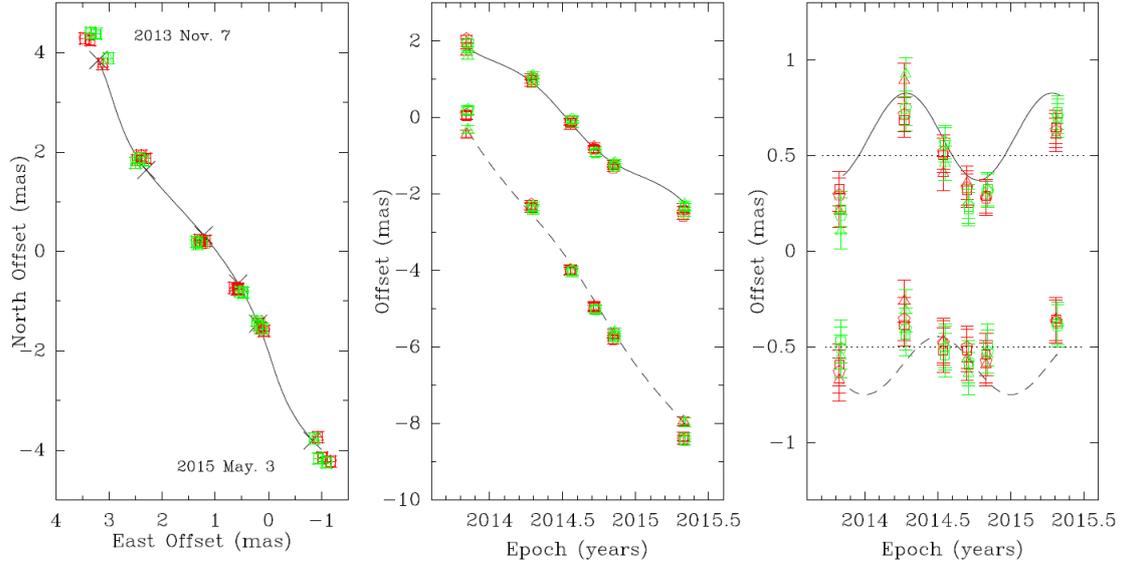

**fig. S1. Parallax and proper motion data and fits for G054.10−00.08.** The position offsets are shown for one maser spot at $V = 37.7$ (triangles) and two maser spots at 36.2 km s$^{-1}$ (squares for the strong spot and hexagons for the weak spot) for the two background sources J1928+1859 (red) and J1935+2031 (green), respectively. *Left Panel:* Positions in the sky, with the first and last epochs labeled. The expected positions from the parallax and proper motion fit are indicated (crosses). *Middle Panel:* East (solid line) and north (dashed line) position offsets and best parallax and proper motions fits versus time. *Right Panel:* Same as the *middle panel*, but with the best-fit proper motion removed for clearer illustration of the parallax sinusoid.

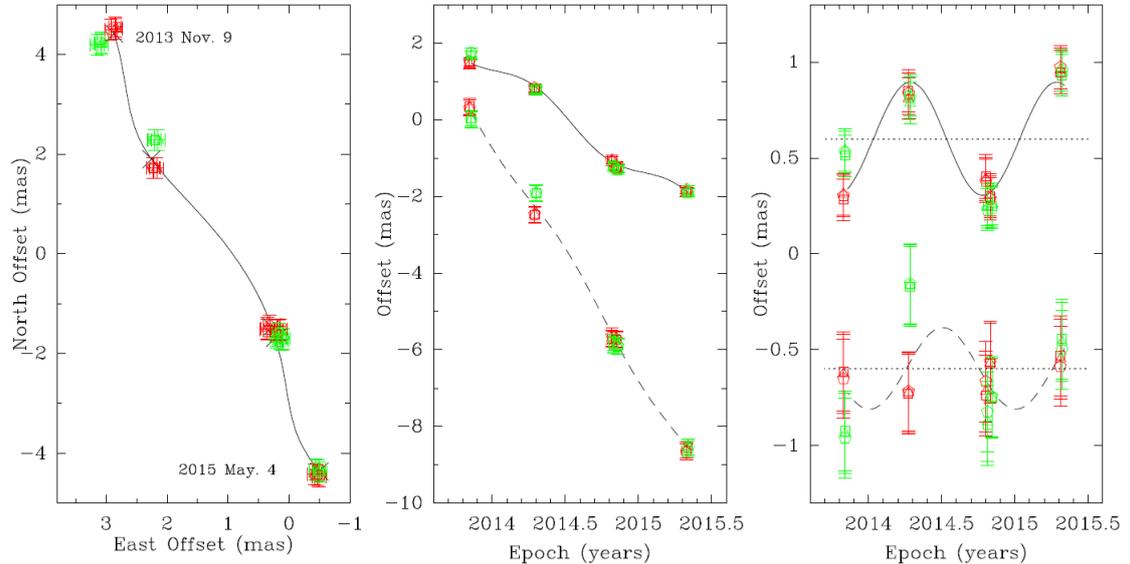

**fig. S2. Parallax and proper motion data and fits for G058.77+00.64.** Position offsets are shown for three maser spots at $V = 33.4$ (triangles), 34.8 (square), and 35.5 (pentagons) km s$^{-1}$ relative to the two background sources: J1936+2357 (red) and J1936+2246 (green), respectively. Solid and dashed lines in the panels represent the same fits as in the corresponding panels of fig. S1.

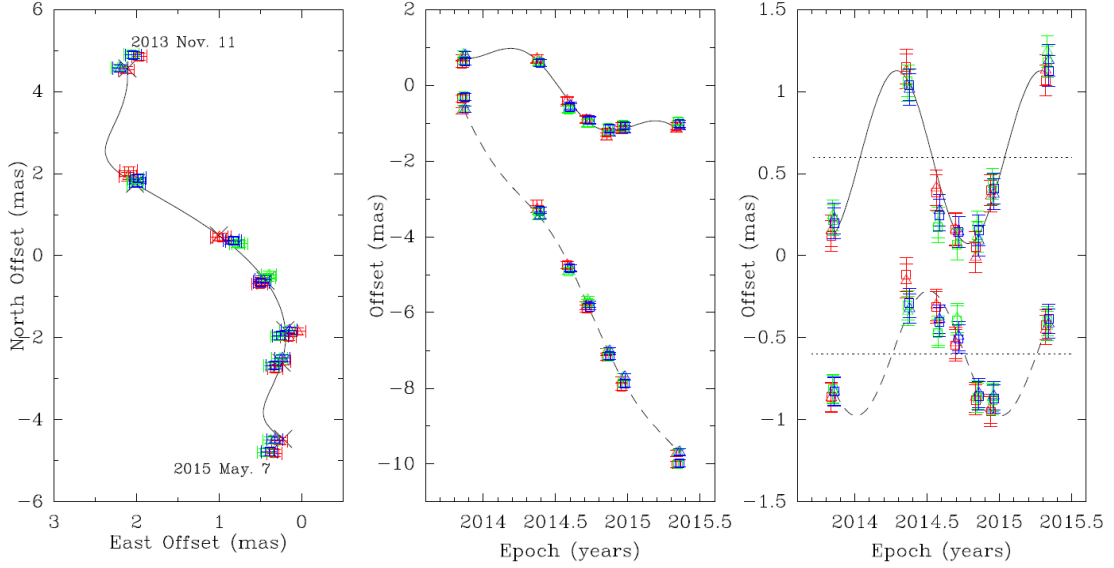

**fig. S3. Parallax and proper motion data and fits for G059.47−00.18.** Position offsets are shown for two maser spots at $V$ = 23.1 (triangles) and 24.6 (square) km s$^{-1}$ relative to the three background sources: J1941+2307 (red), J1946+2300 (green), and J1946+2255 (blue), respectively. Solid and dashed lines in the panels represent the same fits as in the corresponding panels of fig. S1.

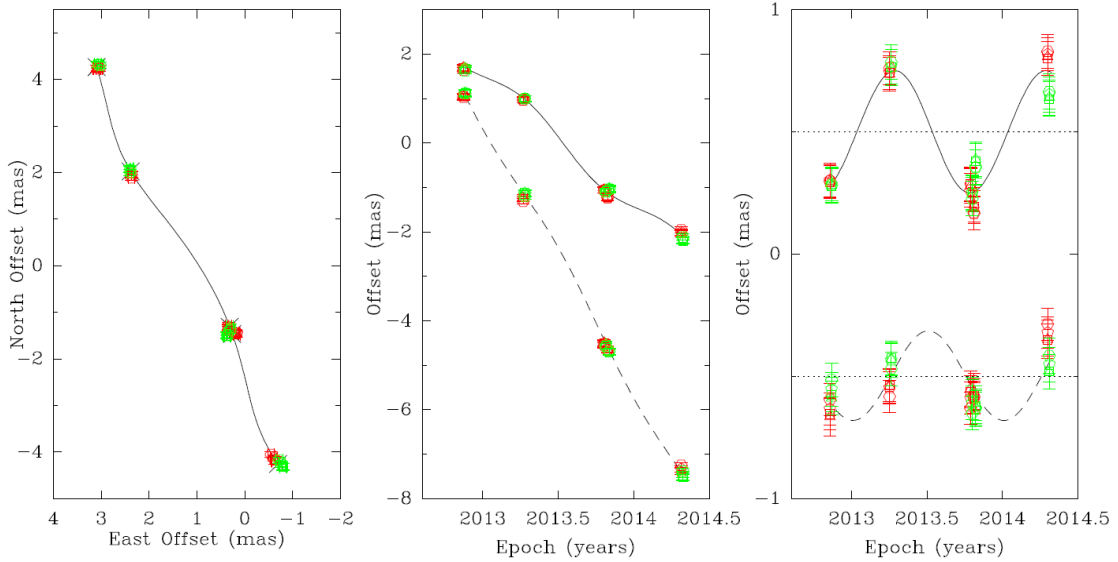

**fig. S4. Parallax and proper motion data and fits for G059.83+00.67.** Position offsets are shown for four maser spots at $V$ = 38.7 (triangles), 38.4 (square), 38.0 (pentagons) and 37.6 km s$^{-1}$ (hexagons) relative to the two background sources J1936+2357 (red) and J1946+2418 (blue), respectively. Solid and dashed lines in the panels represent the same fits as in the corresponding panels of fig. S1.

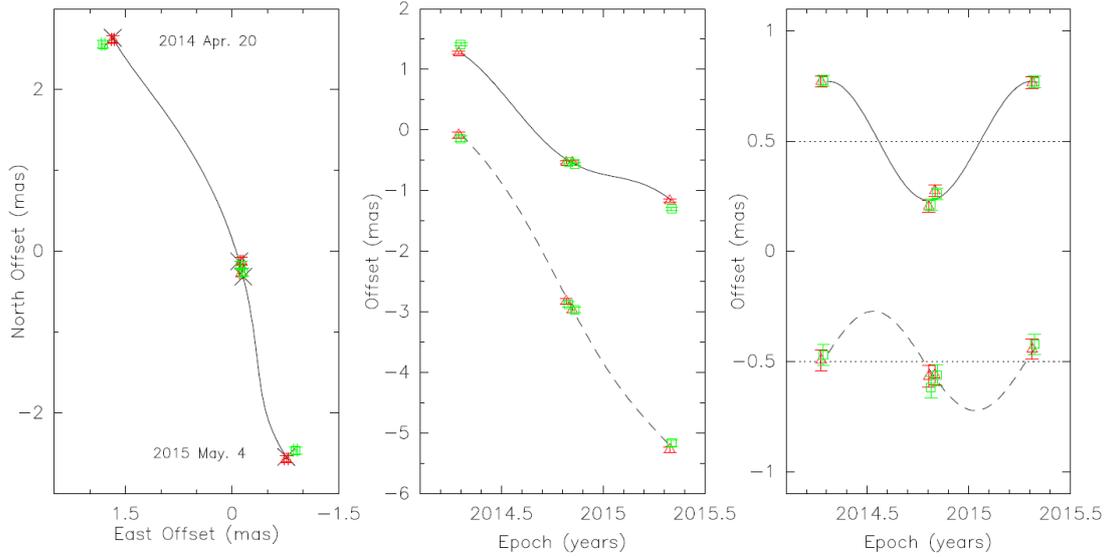

**fig. S5. Parallax and proper motion data and fits for G071.52−00.38.** Position offsets are shown for two maser spots at $V = 10.1$ (red triangles) and 11.0 (green square) km s$^{-1}$ relative to the background source J2015+3410. Solid and dashed lines in the panels represent the same fits as in the corresponding panels of fig. S1.

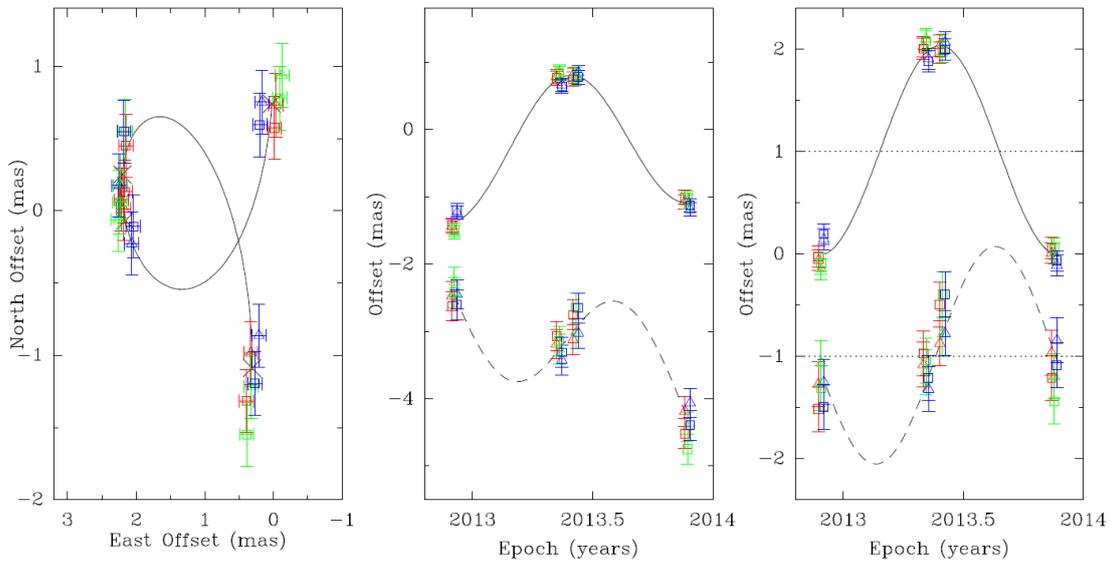

**fig. S6. Parallax and proper motion data and fits for G108.18+05.51.** Position offsets are shown for two maser spots at $V$ = -12.8 (triangles) and -13.2 (square) km s$^{-1}$ relative to the three background sources: J2223+6249 (red), J2231+6218 (green), and J2237+6408 (blue). Solid and dashed lines in the three panels represent the same fits as in the corresponding panels of fig. S1.

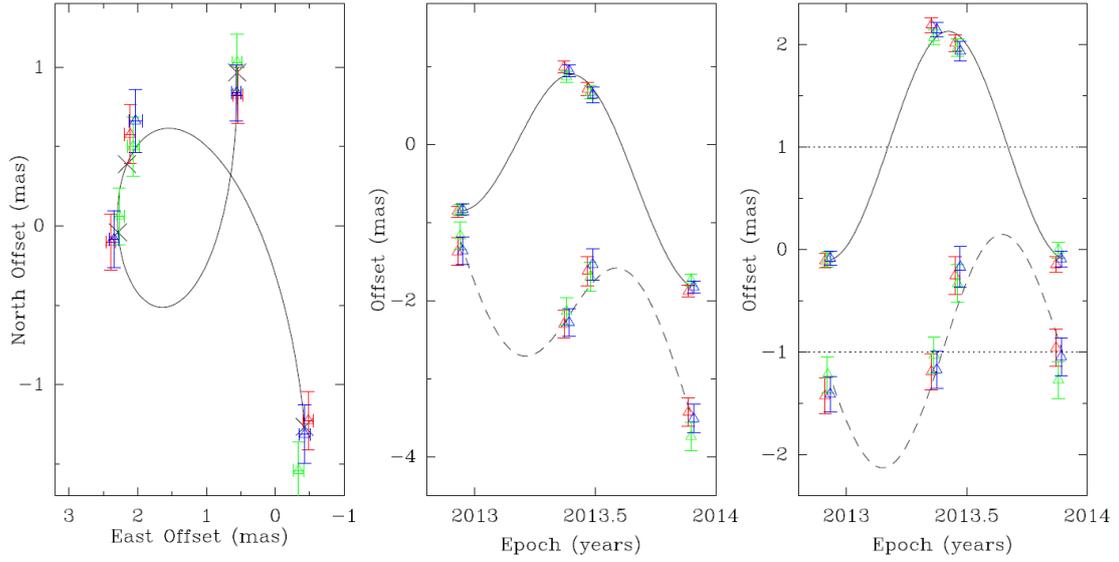

**fig. S7. Parallax and proper motion data and fits for G109.87+02.11.** Position offsets are shown for one maser spot at $V = -2.2$ (triangles) km s$^{-1}$ relative to the three background sources: J2302+6405 (red), J2243+6055 (green), and J2254+6209 (blue). Solid and dashed lines in the panels represent the same fits as in the corresponding panels of fig. S1.

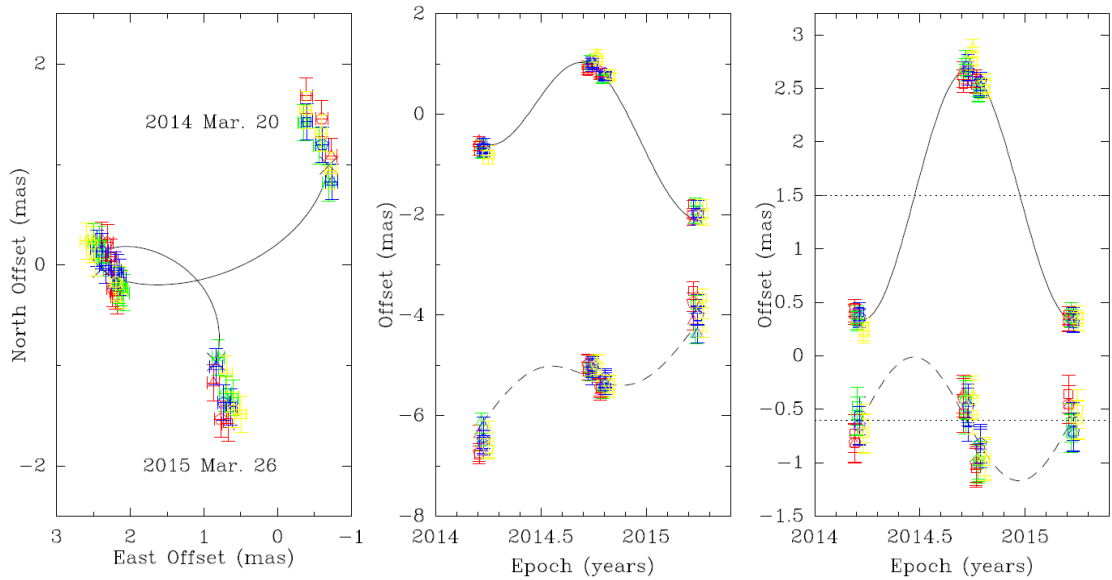

**fig. S8. Parallax and proper motion data and fits for G213.70−12.60.** Position offsets are shown for three maser spots at $V = 10.1$ (triangles), 12.3 (squares), and 13.4 (pentagons) km s$^{-1}$ relative to the four background sources: J0603-0645 (red), J0606-0724 (green), J0605-0759 (blue), and J0607-0834 (yellow). Solid and dashed lines in the panels represent the same fits as in the corresponding panels of fig. S1.